\documentclass[12pt]{article}

\usepackage{amssymb,amstext,amsmath,amsthm}
\usepackage[dvips]{graphicx}
\usepackage{latexsym}
\usepackage{psfrag}
\usepackage{amsfonts}
\usepackage{bbm}

\newcommand{\Ref}[1]{(\ref{#1})}

\newcommand{\eqa}{\begin{eqnarray}}
\newcommand{\neqa}{\end{eqnarray}}
\newcommand{\equ}{\begin{equation}}
\newcommand{\nequ}{\end{equation}}
\newcommand{\no}{\nonumber\\}

\def\om{\omega}
\def\w{\wedge}
\def\la{\langle}
\def\ra{\rangle}
\newcommand{\bra}[1]{\la {#1}|}
\newcommand{\ket}[1]{|{#1}\ra}
\newcommand{\mean}[1]{\la{#1}\ra}

\newcommand{\p}{\partial}
\newcommand{\n}{\hat{n}}

\newcommand{\I}{{\cal I}}
\def\arr{\rightarrow}

\def\ga{\gamma}
\def\th{\theta}
\def\d{\delta}
\def\f{\frac}
\def\wtl{\widetilde}

\usepackage{bbm}
\newcommand{\scr}{\rm\scriptscriptstyle}
\newcommand{\lalg}[1]{\mathfrak{#1}}  
\newcommand{\SU}{\mathrm{SU}}
\newcommand{\U}{\mathrm{U}}
\newcommand{\su}{\lalg{su}}

\newcommand{\spin}{\lalg{spin}}
\newcommand{\Spin}{\mathrm{Spin}}
\let\eps=\epsilon

\def\H{{\cal H}}
\def\si{\sigma}
\def\ss{{\cal S}}

\begin{document}

\title{\Large\bf A new spinfoam vertex for quantum gravity}
\author{{Etera R. Livine$^a$ and Simone Speziale$^b$\footnote{etera.livine@ens-lyon.fr, sspeziale@perimeterinstitute.ca}} \\
\small{${}^a$\emph{Laboratoire de Physique, ENS Lyon, 
46 All\'ee d'Italie, 69364 Lyon, France}}
\\
\small{${}^b$\emph{Perimeter Institute, 31 Caroline St. N, Waterloo, ON N2L 2Y5, Canada}}}
\date{\small\today}

\maketitle

\begin{abstract}
\noindent We introduce a new spinfoam vertex to be used in models of 4d quantum gravity
based on SU(2) and SO(4) BF theory plus constraints.
It can be seen as the conventional vertex of SU(2) BF theory,
the 15j symbol, in a particular basis constructed using SU(2) coherent states.
This basis makes the geometric interpretation of the variables transparent:
they are the vectors normal to the triangles within each tetrahedron.
We study the condition under which these states can be considered semiclassical,
and we show that the semiclassical ones dominate the evaluation of quantum correlations.
Finally, we describe how the constraints reducing BF to gravity can be directly
written in terms of the new variables, and how the semiclassicality of the states
might improve understanding the correct way to implement the constraints.
\end{abstract}

\section{Introduction}
The spinfoam formalism for loop quantum gravity (LQG) \cite{book} is a covariant approach to the
definition of the dynamics of quantum General Relativity (GR).
It provides transition amplitudes between spin network states. The most studied example in the
literature is the Barrett--Crane (BC) model \cite{BarrettC}. This model has interesting aspects,
such as the inclusion of Regge calculus in a precise way, but it can not be considered a complete
proposal. In particular, recent developments on the semiclassical limit show that it does not give
the full correct dynamics for the free graviton propagator
\cite{grav2}. In this paper we introduce a new model that can be taken as the starting point for
the definition of a better behaved dynamics.

Most spinfoam models, including BC, are based on BF theory, a topological theory whose relevance
for 4d quantum gravity has long been conjectured \cite{carloarea}, and has been exploited in a
number of ways. In constructing a specific model for quantum gravity, there is a key difficulty of
quantum BF theory that has to be overcome: not all the variables describing a classical geometry
turn out to commute. This fact has two important consequences. The first is that there is in
general no classical geometry associated to the spin network in the boundary of the spinfoam. This
leads immediately to the problem of finding semiclassical quantum states that approximate a given
classical geometry, in the sense in which wave packets or coherent states approximate classical
configurations in ordinary quantum theory. This is the problem of defining ``coherent states'' for
LQG, which has raised an increasing interest over the last few years \cite{Bombelli, semi}. The second
consequence concerns the definition of the dynamics. This is typically obtained by constraining the BF
theory, a mechanism well understood classically, but still unsettled at the quantum level. The
constraints involve non-commuting variables, and the way to properly impose them is still an open
issue. In particular, it can be argued that the specific procedure leading to the
BC model imposes them too strongly, a fact which also complicates a proper match with
the states of the canonical theory (LQG) living on the boundary. For a recent discussion of this, see \cite{falcao}.
Notice that these key difficulties are present for both lorentzian and euclidean signatures.

Here we consider a definition of the partition function of $\SU(2)$ BF theory
on a Regge triangulation (for a review see for instance \cite{book}),
where the dynamical variables entering the sum have a clear semiclassical interpretation:
they are the normal vectors $\vec n_{t,\tau}$ associated to triangles $t$ within a tetrahedron
$\tau$. The classical geometry
of this discrete manifold (areas, angles, volumes, etc.) can be
described in a transparent way in these variables, provided they satisfy a constraint
for each tetrahedron of the triangulation.
This constraint is the closure condition, that says that the sum
of the four normals associated to each tetrahedron must vanish.
If this condition is satisfied, this new choice of variables positively addresses the issues described above,
in the following way.
First of all, from the boundary point of view, each tetrahedron corresponds to a node of the
boundary spin network. As we show below,
the states associated to the new variables are linear superpositions
of the conventional ones, with the property of minimizing the uncertainty of the non commuting
operators: they thus provide a solution to the problem of finding coherent states.
Upon  satisfying the closure condition, the states are coherent and carry a given classical geometry.
Secondly, as far as the dynamics is concerned, the constraints reducing BF theory to GR
are functions of the full bivectorial structure of the $B$ field.
In our model this structure is related to the vectors $\vec n_{t,\tau}$ in a precise way.
We describe how this improves the construction of the constraints
at the quantum level. We postpone a more detailed analysis of the constraints for further work.

A key point of this approach is the closure condition. As mentioned above, this is crucial for the
semiclassical interpretation of the states. This is not satisfied by all the states entering the
partition function. Yet one of the main results of this paper is to show that quantum correlations
are dominated by the semiclassical states in the large spin limit. The key to this mechanism lies
in the fact that the coherent states we introduce are not normalized, and the configurations
maximizing the norm are the ones satisfying the closure condition. To prove this, we write the norm
as an integral over $\SU(2)$, which we are able to solve explicitly in particular cases. In the
general case, we evaluate the integral using the saddle point approximation, and show that the
configurations satisfying the closure condition are exponentially dominating. All our calculations
are supported by numerical simulations, some of which are reported in the Appendix. Remarkably, the
approximation is accurate even for small spins, and agrees to three
digits with the numerics at spins of order 100.

The results we obtain are valid for $\SU(2)$ BF theory, but can be straightforwardly extended to
$\Spin(4)$, and the same logic applied to any Lie group, including noncompact cases which are of
interest for lorentzian signatures. We postpone such developments for further work, once the
approach considered here proves useful to solve the problems described above.

This paper is organized as follows. In Section \ref{SecVertex}, we introduce the new
vertex amplitude, and describe how it can be obtained following the conventional spinfoam
quantization of BF theory. The new variables are normal vectors associated to the triangles, and
the new vertex built out of coherent intertwiners. In Section \ref{SecCI} we study the norm of
the coherent intertwiners, and show that it is exponentially bigger when the closure condition
holds, thus making semiclassical states dominate the partition function. In Section \ref{SecTet} we
describe the coherent tetrahedron in the conventional basis. In Section \ref{SecQG} we
discuss how the new vertex can be used as a starting point for models of quantum gravity.
We summarize our results in Section \ref{SecConcl}.

Throughout the paper we use unit $\hbar = G = 1$.

\section{The new vertex amplitude}\label{SecVertex}

In this Section we describe the vertex amplitude that characterizes our spinfoam model.
The vertex can be constructed using the conventional procedure for the spinfoam quantization of
$\SU(2)$ BF theory. We refer to the literature for details (see for instance \cite{book}),
and focus here only on the aspect which is relevant to construct the new vertex.

Recall that there is a vector space $\H_0:={\rm Inv}\Big[\bigotimes_{i=1}^F \H_{j_i}\Big]$,
on which $\big(\sum_i \vec{J}_i\big)^2 \equiv 0$, associated with each edge of the spinfoam.
Here the half-integer (spin) $j$ labels the irreducible representations (irreps) of $\SU(2)$, and
$F$ is the number of faces around the edge. If the spinfoam is defined on a Regge triangulation,
$F\equiv 4$ for every edge, yet the considerations of this paper apply to any value of $F$.
The spinfoam quantization assigns
to each edge an integral over the group, that is evaluated by inserting in $\H_0$ the
resolution of identity
\equ\label{ide1}
{\mathbbm 1}_{\H_0} = \sum_{i_1\ldots i_{F-3}} \ket{j_1 \ldots j_F, i_1 \ldots i_{F-3}}
\bra{j_1 \ldots j_F, i_1 \ldots i_{F-3}}.
\nequ
The labels $i$ are called intertwiners, and the sums run over all
half-integer values allowed by the Clebsch-Gordan conditions.
Taking into account the combinatorial structure of the Regge triangulation
(four triangles $t$ in every tetrahedron $\tau$, five tetrahedra in every 4-simplex
$\si$), one ends up with the partition function,
\equ\label{Z}
Z = \sum_{\{j_t\}}\sum_{\{i_\tau\}} \; \prod_t d_{j_t} \; \prod_\si A_\si(j_t, i_\tau),
\nequ
where $d_j = 2j+1$ is the dimension of the irrep $j$, and
the vertex amplitude is $A_\si(j_t, i_\tau) = \{15j\}$, a well known object
from the recoupling theory of $\SU(2)$.

This partition function endows each 4-simplex with 15 quantum numbers, the ten irrep labels $j_t$
and the five intertwiner labels $i_\tau$. Using LQG operators associated with the boundary of the
4-simplex, the ten $j_t$ can be interpreted as areas of the ten triangles in the 4-simplex while the
five $i_\tau$ give 3d dihedral angles between triangles (one -- out of the possible six -- for each
tetrahedron). On the other hand, the complete characterization of
a classical geometry on the 3d boundary requires 20 parameters,
such as the ten areas plus two dihedral angles for each tetrahedron (see next Section for more
details). Therefore the quantum numbers of the partition function are not enough to characterize a
classical geometry.

To overcome these difficulties, we propose to write the same partition function in different variables,
which will endow each 4-simplex with enough geometric information.

The key observation is that $\SU(2)$ coherent states $\ket{j, \n}$ (here $\n$ is a unit vector on
the two-sphere $\ss^2$ -- see below and Appendix \ref{AppCS} for details) provide a (overcomplete) basis
for the irreps. Given an irrep $j$, the resolution of the identity in this representation can
indeed
 be written as ${\mathbbm 1}_j= d_j \int d^2\n \,\ket{j, \n}\bra{j, \n}$, where $d^2\n$ is the
normalized measure on $\ss^2$. By group averaging the tensor product of $F$ coherent states,
we obtain a vector
$$
\ket{\underline{j}, \underline{\n} \, }_0 := \int_{\SU(2)} dh \,\otimes_{i=1}^F h \ket{j_i, \n_i}
$$ in $\H_0$. Here we denoted $\underline{j}$ and $\underline{\n}$ the collections of all
$j_i$'s and $\n_i$'s.
The set of all these vectors
when varying the $\n_i$'s forms an overcomplete basis in ${\H}_0$. Using this basis the
resolution of the identity in $\H_0$ can be written as
\eqa\label{ide2}
{\mathbbm 1}_{\H_0}  &=&
\int \prod_{i=1}^F d^2\n_i \ d_{j_i} \ 
\ket{\underline{j}, \underline{\n} \, }_0\bra{\underline{j}, \underline{\n} \, }_0 = \no
&=& \int \prod_{i=1}^F d^2\n_i \ d_{j_i}
\int dh \int  dh' \; h \ket{j_i, \n_i}\bra{j_i, \n_i} h'.
\neqa

This formula can be used in the edge integrations instead of \Ref{ide1}. The combinatorics is the
same as before. In particular, notice that \Ref{ide2} assigns to a triangle
$t$ a normal $\n_t(\tau)$ for each tetrahedron sharing $t$. Within a single 4-simplex, there
are two tetrahedra sharing a triangle, which we denote $u(t)$ and $d(t)$.
Furthermore, \Ref{ide2} assigns two group integrals
to each tetrahedron, one for each 4-simplex sharing it. Taking also into account the $d_j$ factors
in \Ref{ide2} and using the conventional spinfoam procedure, one gets
\equ\label{Z2}
Z= \sum_{\{j_t \}} \int \prod_{t,\tau} d^2\n_{t,\tau} \; \prod_t d_{j_t}{}^3 \;
\prod_\si \, A_\si(j_t, \n_{t,\eps(t)})
\nequ
where the new vertex is
\equ\label{vertex}
A_\si = \int \prod_\tau dh_\tau \, \prod_t
\, \bra{j_t, \n_{t, u(t)}} h_{u(t)}^{-1} h_{d(t)} \ket{j_l, \n_{t, d(t)}}.
\nequ

Let us add a few comments.
\begin{itemize}
\item
The new vertex gives a quantization of BF theory in terms of the variables $j_t$ and
$\n_{t,\tau(t)}$. Together, these represent the full bivector $B^i_{\mu\nu}(x)$ discretized on triangles
belonging to tetrahedra. Taking the $B$ field constant on each tetrahedron, the discretization
and quantization procedures can be schematically summarized as follows,
\equ\label{B}
B^i_{\mu\nu}(x) \ \stackrel{\rm discr}{\longrightarrow} \
B_t^i(\tau) := \int_t B^i_{\mu\nu}(x) \, dx^\mu \w dx^\nu
\ \stackrel{\rm quant}{\longrightarrow} \ j_t \, \n_{t,\tau}.
\nequ
In the quantum theory, the field $B^i_{\mu\nu}(x)$ is represented by a vector
$j_t \, \n_{t,\tau(t)}$ associated to each triangle in a given tetrahedron. The set of all these vectors
can be used to describe a classical discrete geometry.

On the other hand, the conventional quantization of BF theory differs in the quantization step, which reads
\equ\label{BJ}
B_t^i \ \stackrel{\rm quant}{\longrightarrow} \ J^i_t,
\nequ
where $\vec J_t$ are $\SU(2)$ generators associated to each triangle. Consequently the variables
entering the partition function \Ref{Z} are irrep labels $j_t$ and intertwiner labels $i_\tau$.
The first variables are related to discretization of the modulus of $B$, and thus to the
area of triangles. The intertwiner labels are related to \emph{one} angle (out of six)
for each tetrahedron. Therefore in these variables a classical geometry is hidden, and this in turn makes it
hard to implement the dynamics, as we discuss below in Section \ref{SecQG}.

\item
This vertex can be used directly for $\SU(2)$ BF theory, and straightforwardly generalized to the
$\Spin(4)$ case, exploiting the homomorphism $\Spin(4) = \SU(2)\times \SU(2)$. In particular, \Ref{B} becomes
\equ\label{B1}
B^{IJ}_{\mu\nu}(x) \ \stackrel{\rm discr}{\longrightarrow} \
B_t^{IJ}(\tau) := \int_t B^{IJ}_{\mu\nu}(x) \, dx^\mu \w dx^\nu
\ \stackrel{\rm quant}{\longrightarrow} \
j^+_t \, \n^+_{t,\tau} \oplus j^-_t \, \n^-_{t,\tau}.
\nequ

\item
There is a simple relation between this vertex and the one given in \Ref{Z} which is the one
usually used. Since all we did is simply a change of basis, \Ref{vertex} can be written
as a linear combination of $\{15j\}$'s as we will see in more details in Section \ref{SecTet}.

\item

The BC vertex can be obtained from \Ref{vertex}: performing the $d^2\n_i$ integrations
with the condition $\n_{t,u(t)} \equiv \n_{t,d(t)}$
(namely inserting terms $\d(\n_{t,u(t)}-\n_{t,d(t)})$ in \Ref{Z2}), we indeed obtain the BC vertex in its integral
representation \cite{Barrett},
\equ
A_{\rm BC} =
\int_{\SU(2)} \prod_\tau dh_\tau \, \prod_t \f1{d_{j_t}} \ \chi^{(j_t)}(h_{u(t)}^{-1} h_{d(t)}).
\nequ
\end{itemize}

To understand the geometry of this new vertex amplitude, it is crucial to study its
asymptotic behaviour in the large spin
limit.\footnote{The reasons why the large spin limit is relevant to discuss the dynamics
of spinfoams is largely discussed in the literature (see for instance \cite{book}). The
main reason is that this limit appears to be related to Regge calculus, a discrete version
to classical GR.}
In the rest of this paper, we begin the analysis of this limit, by studying in
details the behavior of the states $\ket{\underline{j}, \underline{\n} \, }_0$ entering \Ref{vertex}.
The full asymptotics of the vertex will be reported elsewhere. Nonetheless,
it will become clear from the rest of the paper that the large spin limit of this
integral is dominated by values of the group elements such that the factors of the
integrand in \Ref{vertex} are close to one, namely such that
\equ\label{normals}
\ket{j_t, \n_{t,u(t)}} = h_{u(t)}^{-1} h_{d(t)} \ket{j_t, \n_{t,d(t)}}.
\nequ
This has a compelling geometric interpretation. Recall that $\n_{t,u(t)}$ and $\n_{t,d(t)}$ are the normals
to the same triangle as as seen from the two tetrahedra sharing it. Then \Ref{normals} means that they
are changed into one another by a gravitational holonomy.

This is in contrast with the BC model, which fixes $\n_{t,u(t)} \equiv \n_{t,d(t)}$,
thus not allowing any gravitational parallel transport between tetrahedra.
This is another way of seeing the well known problem that the BC model has not enough
degrees of freedom.
We will come back to this point in Section \ref{SecQG}, where we discuss how this vertex can be taken as
the starting point for quantum gravity models.

\section{Coherent intertwiners}\label{SecCI}
In this Section we focus on the building blocks of the new vertex, namely the states $\ket{\underline{j}, \underline{\n} \, }_0$
associated to the tetrahedra entering \Ref{ide2}.
Before studying the details of the mathematical structure of these states, let us discuss their physical
meaning. In the canonical picture, a tetrahedron is dual to a 4-valent node in a spin network.
More in general, when the edge of the spinfoam is on the boundary, there is a one--to--one
correspondence between the number of faces $F$ around it, and the valence $V$ of the node
of the boundary spin network.
Given a discrete atom of space dual to the node with $V$ faces, its classical geometry can be entirely determined
using the $V$ areas and $2(V-3)$ angles between them.
Alternatively, one can use the (non--unit) normal vectors $\vec n_i$ associated with the faces,
constrained to \emph{close}, namely to satisfy $\sum_i \vec n_i =0$.

On the other hand, the conventional basis used in \Ref{ide1} gives quantum numbers for $V$ areas
but only $V-3$ angles. This is immediate from the fact that one associates $\SU(2)$ generators
$\vec J_i$ with each face \cite{Barbieri}, and only $V-3$ of the possible scalar products $\vec J_i \cdot \vec J_k$
commute among each other. Thus half the classical angles are missing.
To solve this problem,
we argue that the states $\ket{\underline{j}, \underline{\n} \, }_0$ carry enough information to describe a classical geometry
for the discrete atom of space dual to the node. In particular, we interpret
the vectors $j_i \, \n_i$ precisely with the meaning of
normal vectors $\vec n_i$ to the triangle, and we read the geometry off them
as described above. This requires the vectors to satisfy the following constraint:
\equ\label{closure}
\vec N = \sum_{i=1}^V j_i \, \n_i = 0.
\nequ
This \emph{closure condition} will be studied in detail in the rest of this Section.
If \Ref{closure} holds, all classical geometric observables can be parametrized as
${\cal O}(\{j_i\,\n_i \})$.

This is the physical picture we consider. To describe the mathematical details,
let us recall basic properties of the $\SU(2)$ coherent states (for details see the Appendix \ref{AppCS}).
A coherent state is obtained via the group action on the highest weight state,
\equ\label{CS}
\ket{j,\n} =g(\hat{n})\ket{j,j},
\nequ
where $g(\hat{n})$ is a group element rotating the north pole $\hat{z}\equiv(0,0,1)$ to the
unit vector $\hat{n}$ (and such that its rotation axis is orthogonal to $\hat{z}$).
These coherent states are semiclassical in the sense that
they localize the direction $\n$ of the angular momentum.
We have $\bra{j, \n}\vec J \, \ket{j, \n}=j \, \n$ and relative uncertainty
$\Delta\sim1/\sqrt{j}$ vanishing in the large spin limit (see Appendix \ref{formulas}).
As $\ket{j,j}$ has direction $z$ with minimal uncertainty, $\ket{j, \n}$ has direction $\n$
with minimal uncertainty.

This localization property is preserved by the tensor product of irreps $\otimes_i \ket{j_i, \n_i}$:
in the large spin limit we have for instance 
$\mean{\vec J_i \cdot \vec J_k} \simeq j_i \, j_k\, \n_i\cdot \n_k$ for any $i$ and $k$.
More in general any operator $\hat{\cal O}(\{\vec J_i \, \})$
on the tensor product of coherent states satisfies
\equ\label{ev}
\mean{\hat{\cal O}(\{\vec J_i \, \})} \simeq {\cal O}(\{j_i\,\n_i \})
\nequ with vanishing relative uncertainty.

So far, the $\n_i$'s are completely free parameters. If
the closure condition \Ref{closure} holds, \Ref{ev} tells us that the states admit a semiclassical
interpretation.
Consequently, states in $\bigotimes_{i=1}^V \H_{j_i}$ satisfying \Ref{closure} can
be considered semiclassical. Yet they cannot be used in the partition function:
the tensor product of coherent states must be projected on the
invariant subspace $\H_0$, in order to implement gauge invariance.
This can be achieved by group averaging, as anticipated in the previous
Section, and the resulting states
written as a linear combination of the conventional basis of intertwiners are
\equ\label{n0}
\H_0 \ni \ \ket{\underline{j}, \underline{\n} \, }_0 := \int dh \,\otimes_{i=1}^V h \ket{j_i, \n_i} =
  \sum_{i_1\ldots i_{V-3}}
c_{i_1\ldots i_{V-3}}(j_i, \n_i) \ \ket{j_1 \ldots j_V, i_1 \ldots i_{V-3}}.
\nequ
We call these states coherent intertwiners.
The explicit form of the $c_{i_1\ldots i_{V-3}}$ coefficients is reported in the Appendix
\ref{AppCS}, but it will not be relevant for the rest of the paper. In Section \ref{SecTet}
below we will study their asymptotics in the 4-valent case.

Crucially, \Ref{ev} holds on the states $\ket{\underline{j}, \underline{\n} \, }_0$ (see Appendix \ref{formulas}), thus
group averaging does not spoil the semiclassical interpretation of these states in the large spin
limit. But what is more important, group averaging correlates the $\n_i$'s, making the state
sensitive to \Ref{closure}. Indeed, notice that a state in $\H_0$ satisfies $\big{(}\sum_i \vec
J_i\big{)}^2=0$ but not necessarily the closure condition. The main result of this paper is to show
that the states satisfying \Ref{closure} have an exponentially bigger norm in the large spin limit.
Therefore every state \Ref{n0} enters the partition function \Ref{Z2}, but quantum correlations
will be dominated by semiclassical states in the large spin limit. We can then say that projecting
onto $\H_0$ does not force the $V$-simplex to close identically, but it does imply that closed
simplices dominate the dynamics.

\subsection{The closure condition}

The key for the mechanism is the following: the coherent interwiner states \Ref{n0} are not
normalized, and the norm is a function of the $\n_i$,
\equ\label{norm}
f(\n_i) := {}_0\la{\underline{j}, \underline{\n} \,}\ket{\underline{j}, \underline{\n} \, }_0
= \int dh \, \bra{j_i, \n_i} h^{\otimes V} \ket{j_i, \n_i} \neq 1.
\nequ
Consequently, the relative weight of these states will depend on the configuration of the $\n_i$.
As we show below, the norm is exponentially smaller when the closure condition \Ref{closure}
is not satisfied.

To give a simple picture of the way this happens, let us first consider the
trivial case of a bivalent node. In this case, the norm \Ref{norm} can be straightforwardly
evaluated using the explicit values of the Clebsch-Gordan coefficients.
Using $\ket{j_i, \n_i}= g_i \ket{j_i,j_i}$, we can write
\equ\label{normrep}
\bra{j_i, \n_i}h \ket{j_i, \n_i} = D^{(j_i)}_{j_i k}(g_i{}^{-1})\,D^{(j_i)}_{kl}(h) \, D^{(j_i)}_{lj_i}(g_i).
\nequ
The integration over $h$ in \Ref{norm} gives the
orthogonality condition for the Clebsch-Gordan coefficients,
\equ\label{CBorto}
\int dh \, D^{(j_1)}_{kl}(h) \, D^{(j_2)}_{mn}(h) =
\f{\d^{j_1\,j_2}}{d_{j_1}} \, \d_{k, -m} \, \d_{l, -n},
\nequ
and we get $f(\n_1, \n_2) = \f{\d^{j_1\,j_2}}{d_{j_1}} \, \Big| D^{(j_1)}_{j_1,-j_1}(g_1^{-1} g_2) \Big|^2$.
This matrix element can be evaluated starting from the fundamental representation.
As we show in the Appendix \ref{formulas}, we have
\equ\label{abidal}
\Big| D^{(j_1)}_{j_1,-j_1}(g_1^{-1} g_2) \Big|^2 =
\Big|   \bra{+} g_1^{-1} g_2 \ket{-} \Big|^{4j_1} = \Big(1-\f14\, (\n_1+\n_2)^2\Big)^{2j_1}.
\nequ
Defining $J= \sum_i j_i$, we conclude that
\equ\label{result}
f(\n_1, \n_2) = \f{\d^{j_1\,j_2}}{d_{j_1}} \,
\left(1-\f{\vec N^2}{J^2} \right)^{2j_1}.
\nequ
As $\vec N^2 \leq J^2$, we see that in the large spin limit the norm vanishes exponentially, unless the closure
condition is satisfied, in which case it vanishes with an inverse power law.
Consequently, the norm of the states satisfying the closure condition becomes exponentially
bigger than the others.

Notice that in the bivalent case, the closure condition means that the two vectors are
antiparallel. The situation is analogous for the trivalent case: closed configurations will
correspond to three coplanar vectors. It is only for valence four or higher that the geometry of
the problem becomes more interesting, and one has closed configurations spanning 3d space. However,
the configurations where the vectors are all aligned are still present. Since such configurations
fail to produce 3d structures, we refer to them as degenerate.

For nodes with higher valence we expect a result analogous to \Ref{result}. However
the analysis is far more intricate, the reason being that now when we use
\Ref{normrep} we obtain an integral $\int dh \, \prod_i D^{(j_i)}_{m_in_i}(h)$ of $V$
representation matrices, and in general we have no simple formula for the (generalized)
Clebsch-Gordan coefficients (see Appendix \ref{AppCS}) appearing from this integration. It is
then simpler to study the norm explicitly using the fact that (see Appendix \ref{formulas})
\equ\label{cis}
\bra{j,j}g(\n)^{-1} \, h \, g(\n)\ket{j,j} =
\Big(\bra{\f12,+} g(\n)^{-1} \, h \, g(\n) \ket{\f12,+}\Big)^{2j}
=(\cos\ga + i \sin\ga \,\hat{u}\cdot \n)^{2j},
\nequ
where we parametrized $\SU(2)$ group elements as
\equ\label{g}
h(\gamma, \hat u) = \cos\gamma \, \mathbbm{1}+ i \, \sin\gamma \, \hat u\cdot \vec\sigma,
\quad \gamma\in[0, \pi], \quad \hat{u}\in \ss^2.
\nequ
It is convenient to introduce the vector $\vec{p}\,=\,\sin\ga \, \hat{u}$. In
these variables, the Haar measure splits into two integrals over the
unit three-ball $B: |\vec p\,|\le 1$, corresponding to the northern
(say $B_+$) and southern (say $B_-$) ``hemispheres'' of $\SU(2)\simeq S^3$,
$$
\int_{\SU(2)} dh
\,=\,\f{1}{2\pi^2}\sum_{\eta=\pm}\int_{B_\eta} \f{d^3 \vec{p}}{\sqrt{1-\vec{p} \,{}^2}}.
$$

We then write \Ref{norm} as
\equ\label{normcos}
f(\n_i) 
= \f1{2\pi^2}\sum_{\eta=\pm} \int_{B_\eta} \f{d^3\vec p}{\sqrt{1-\vec{p} \,{}^2}} \,
\prod_{i=1}^V (p_\eta + i \vec p \cdot \n_i)^{2j_i}.
\nequ
where $p_\eta=\eta\sqrt{1-\vec{p} \,{}^2}$ in $B_{\eta}$. It is a triple integral depending on $3V$
parameters. Notice that the measure tends to locate the integral on the boundary $|\vec p\,|=1$ of
the integration domain, whereas the integrand tends to locate it on the origin $|\vec p\,|=0$. In
the large spin limit, the latter behaviour will of course dominate. Since we are interested in the
asymptotic behavior of this norm, it is enough to study its saddle point approximation in the large
spin limit.
In the Appendix \ref{AppSpher} 
we discuss the exact evaluation of the norm.


\subsection{Saddle point analysis}\label{SecSaddle}
We want to study the behavior of \Ref{normcos} in the large spin limit. To this end, we scale the
spin labels by an overall constant $\lambda$, \emph{i.e.} $j_i\mapsto \lambda \, k_i$ and write
\Ref{normcos} as
\equ\label{action}
f(\n_i) = \f{1}{2\pi^2}\sum_{\eta=\pm}\int_{B_\eta} \f{d^3 \vec{p}}{\sqrt{1-\vec{p} \,{}^2}}
\; e^{\lambda S(\vec p)}, \qquad
S(\vec p\,) := \sum_i 2k_i \, \ln \big(p_\eta + i \vec p \cdot \n_i\big).
\nequ
Notice that the real part of $S$ is negative.
When the variable $\lambda$ is taken to infinity,
the integral can be approximated computing the saddle point expansion and evaluating
the Gaussian integral on the maxima of $S$.\footnotemark
\footnotetext{
Let us point out that the measure divergence at $|p|=1$ does not contribute at all to the leading order behaviour of the norm $f(\hat{n}_i)$ unlike the case of the asymptotics of the 6j and 10j symbols \cite{Freidel}. Indeed, the divergence in $\f1{\sqrt{1-\vec p\,^2}}$ is still integrable and does not lead to a divergent integral.}
It is immediate to see that the norm of $S$ is maximized by $\vec p = 0$, corresponding to
$h=\mathbbm 1$, but the phase requires more attention.
The saddle points of $S(\vec p\,)$ on $B_{\eta}$ satisfy
\equ\label{gradient}
\nabla S =
\sum_i 2k_i \; \f{(\vec{p}\cdot\hat{n}_i) \, \hat{n}_i - \vec{p}}{p_\eta^2+(\vec{p}\cdot\hat{n}_i)^2} +
i\; \f1{p_\eta}\;
\sum_i 2 k_i \; \f{(\vec{p}\cdot\hat{n}_i) \, \vec{p} +
p_\eta^2 \, \hat{n}_i}{p_\eta^2+(\vec{p}\cdot\hat{n}_i)^2} = 0.
\nequ
A solution of this set of six real equations can be found
taking the scalar product of the real part with $\vec{p}$, giving
$$
\sum_i k_i \,\left(1-\f1{p_\eta^2+(\vec{p}\cdot\hat{n}_i)^2}\right) =0.
$$
Since $0\le p_\eta^2 + (\vec{p}\cdot \hat{n}_i)^2 \le1$ and
$k_i>0$ for all $i$'s, this implies that
$p_\eta^2+(\vec{p}\cdot\hat{n}_i)^2=1$ for all $i$, namely
$\vec p \;{}^2 = (\vec{p}\cdot\hat{n}_i)^2$.

This leaves us with two cases:
\begin{itemize}
\item Either there exists (at least) two non-collinear vectors, $\hat{n}_i\ne\pm
\hat{n}_j$, then the latter equality forces $\vec{p}=0$ and the imaginary part of
\Ref{gradient} simply reads $\sum_i k_i \, \hat{n}_i=0$.
\item Or all the $\hat{n}_i$ are equal to the same unit vector
$\hat{n}$ up to a sign, then a little work on the imaginary part of \Ref{gradient} leads us
to the same constraint $\sum_i k_i \, \hat{n}_i=0$ but leaves no
constraint on the vector $\vec{p}$.
\end{itemize}
In both cases, we see that we have derived the closure condition \Ref{closure} from the saddle
point analysis in the large spin limit $\lambda\arr\infty$. Numerical investigations show that
there are no other solutions to \Ref{gradient}, and that there are no saddle points if the closure
condition is not satisfied. Which as we show below means that as $\lambda$ increases the norm
\Ref{norm} is exponentially smaller for non-closed configurations, and thus correlations on spin
networks in the large spin regime will be dominated by those intertwiners satisfying the closure
condition and thus describing a classical geometry.

Let us start by the non-degenerate generic situation with non-collinear $\hat{n}_i$'s. Then $S(0) =
0$ and the Hessian matrix of the second derivatives at the two (one for each three-ball) fixed
points $\vec{p}=0$ is a sum of projectors $P_{ab}^{i} = \, \Big(\d_{ab} - (\n_i)_a (\n_i)_b
\Big)$ orthogonal to the unit vectors $\hat{n}_i$,
\equ\label{H}
H_{ab}\,\equiv \,
-\f12\left.\f{\p^2}{\p p_a\p p_b}\right|_{p=0}\,=\,
\sum_i  \, k_i \, P_{ab}^{i}.
\nequ
Notice that this is independent of $\eta$, thus both saddle points
give equal contributions.
Next, the property $\sum_{i,j} \, k_i \, k_j \, (n_i \cdot n_j)^2 < \sum_{i,j} \, k_i \, k_j$
guarantees that the eigenvalues of $H_{ab}$ are all positive.
Therefore the Gaussian integral on the real line converges and we can straightforwardly
compute the saddle point approximation,
\equ\label{saddleclosed}
f(\n_i)\,=\,
\f{1}{\pi^2}\int d^3\vec{p}\,
\; e^{-{\lambda} \, H_{ab} \, p_a \, p_b}
\,=\, \f{1}{\sqrt\pi}\f{1}{\lambda^{3/2}\sqrt{\det H}}.
\nequ

We can compute $\det H$ explicitly in terms of the labels $k_i$ and normals $\hat{n}_i$:
$$
\det H=\f{K}{2}\sum_{i,j}k_i \, k_j \, |\n_i\w \n_j|^2
-\f16\sum_{i,j,k}k_i \, k_j \, k_k \, |\n_i\cdot(\n_j\w \n_k)|^2,
$$
with $K=\sum_i k_i$. When the closure constraint is satisfied, i.e. $\sum_i k_i\n_i=0$,  the quadratic
term is a function of the (squared) area of the internal parallelograms while the cubic
term relates to the (squared) volume of the tetrahedron. From this point of view, the determinant
$\det H$ measures the shape of the tetrahedron at fixed triangle area $k_i$.

The result \Ref{saddleclosed} has been numerically confirmed for various closed configurations
and different values of $V$. As an example, a plot is reported in Appendix \ref{AppPlots},
which shows explicitly how accurate the approximation is, even for very small spins.
In particular, we have one digit of accuracy from the beginning, and three digits at spins
of order 100.
We conclude that the norm of coherent intertwiners satisfying the closure condition scales
as $\lambda^{-3/2}$ in the large spin limit.

\subsection{Degenerate configurations}\label{SecDeg}
Let us now look at the case where all the $\n_i$'s are collinear.
Without loss of generality, we can take all the $\n_i$'s aligned with the $z$ axis, namely
$\n_i = \eta_i \, \hat{z}$, $\eta_i=\pm$. Let us also define
$j^\pm = \sum_{i | \eta_i=\pm} j_i$.
One can easily convince himself that\,\footnotemark \ $f(\eta_i \, \hat{z})\equiv 0$ unless $\prod_i \eta_i=1$
and $j^+=j^- \equiv \f{V}2 \lambda$.
\footnotetext{It suffices to
decompose the state as in \Ref{n0} and recall the conditions for a tensor
product of states to admit singlet irreps.}
Then, \Ref{cis} reads $(\cos\ga + \eta_i \, \sin\ga \, u_z)^{2j_i}$.
We take the conventional parametrization
$\ss^2\ni \hat u = (\cos\alpha \sin\beta, \sin\alpha \sin\beta, \cos\beta)$,
with normalized measure $d^2\hat u = \f1{4\pi} \, d\alpha \, d\beta \, \sin\beta$. We see that the $\alpha$
angle drops out, and the integral is simply
\equ\label{mb1}
\f1\pi \int_0^\pi d\beta \, d\ga \, \sin\beta \, \sin^2\ga \,
(1-\sin^2\beta \, \sin^2\ga)^{V \lambda}.
\nequ
This integral can be exactly evaluated as shown in the Appendix \ref{AppSpher}, and the result is $(V\lambda+1)^{-1}$.

We conclude that
\equ
f(\lambda, \eta_i \, \hat{z}) =
\left\{\begin{array}{ll}
\f1{V\lambda+1} & {\rm if} \ \prod_i \eta_i=1, \ j^+=j^- \equiv \f{V}2 \lambda, \\ & \\
0 & {\rm otherwise.}
\end{array}\right.
\nequ

Notice that the degenerate configurations scale as $\lambda^{-1}$, whereas the closed ones scale as $\lambda^{-3/2}$.
For $\lambda\mapsto \infty$, the degenerate configurations dominate
as $\sqrt\lambda$. However,
in the partition function \Ref{Z2} they have zero measure, thus we do not expect them to affect
the dynamics in a relevant way. Furthermore these degenerate configurations
are also present in BC, and they do not enter the computation of physically relevant quantities,
as conjectured in \cite{RovelliProp} and shown in \cite{grav3}.

\subsection{Non-closed configurations}\label{SecNC}

For configurations such that $\vec N \neq 0$,
numerical simulations show that there are no saddle points, but the integrand
in \Ref{normcos} is still maximized at $\vec p=0.$ For notational consistency, let us also
rescale $\vec N \mapsto \lambda \vec N$. The expansion of $S$ in \Ref{action}
around $\vec p=0$ reads
\equ\label{expnonclosed}
S(\vec p \, ) = 2 \,i \, \vec N \cdot \vec p - {H}_{ab}\, p_a \, p_b + o(p^3).
\nequ
If we approximate the integral using this expansion for the exponent, we see the presence
of a phase term coming from the linear term in \Ref{expnonclosed}. This phase term dumps the integral exponentially.
In fact, a simple calculation gives
\equ\label{nonclosed}
f(\n_i)\,=\,
\f{1}{\pi^2}\int d^3\vec{p}\,
\; e^{i \, 2\lambda \,\vec N \cdot \vec p \, - {\lambda} \, {H}_{ab} \, p_a \, p_b}
\,=\, \f{1}{\sqrt\pi}\f{1}{\lambda^{3/2}\sqrt{\det {H}}} \,
\left(e^{- \vec N \cdot {H}^{-1} \vec N}\right)^\lambda.
\nequ
This result is again supported by numerical simulations for various non-closed
configurations and different values of $V$ (see Appendix \ref{AppPlots}). As $ \vec N \cdot {H}^{-1} \vec N > 0$, the norm of non--closed configurations is exponentially smaller than the closed ones.

We do not have a simple explicit expression for the inverse Hessian $H^{-1}$, but we can easily
write it as a power series. Notice that $H$ in \Ref{H} reads as $K ({\mathbbm 1}-\Sigma)$ with $K=\sum_i k_i$ and
$\Sigma_{ab}=\sum_i (k_i/K)\, \n_i^a\, \n_i^b$. As the norm of $\Sigma_{ab}$ is smaller than 1, the inverse
series for $H^{-1}$ converges (except in the degenerate configurations where some of the $\n_i$ are
collinear):
$$
H^{-1}\,=\,
\f1K\big{[}{\mathbbm 1}+\Sigma+\Sigma^2+\dots\big{]}
\,=\,
\f1K\big{[}{\mathbbm 1}+\sum_i \f{k_i}{K} \, \n_i^a \, \n_i^b
+\sum_{i,j} \f{k_ik_j}{K^2}\, (\n_i\cdot\n_j)\,\n_i^a \, \n_j^b+\dots
\big{]}.
$$
This allows us to express the exponent of \Ref{nonclosed} 
in terms of the matrix $G_{ij} \equiv \f{\sqrt{k_i \, k_j}}K \, (\n_i\cdot\n_j)$,
\eqa
N\cdot H^{-1}N&=&
\f{1}{K}\sum_{i,j} k_{i} \, k_{j} \, (\n_i\cdot\n_j)
+\f1{K^2}\sum_{i,j,k} k_i \, k_j \, k_k \, (\n_i\cdot\n_j) \, (\n_j\cdot\n_k)
+\dots \no\nonumber
&=& \sqrt{k_i} \, (G+G^2+G^3+\dots)_{ij} \sqrt{k_j}.
\neqa
Notice that $G_{ij}$ is a $V\times V$ Gram matrix of scalar products of unit vectors. For $V$
greater than 3, the vectors $\n_i$ are not linearly independent and $\det G_{ij}=0$. Nevertheless the
series $G+G^2+G^3+\dots$ can be re-summed as $G({\mathbbm 1}-G)^{-1}$.
Thus it appears that the speed of convergence of the intertwiner norm in this non-closed case is
directly related to the eigenvalues of the Gram matrix $G$. More precisely, estimating its
smallest and largest eigenvalues (both between 0 and 1) would allow to bound the convergence speed
of the norm as $\lambda$ goes to infinity.

\section{Four-valent case: the coherent tetrahedron}\label{SecTet}

Above we have constructed coherent intertwiners for nodes of generic valence. The 4-valent case,
whose dual geometric picture is a tetrahedron, is
of particular interest as it enters the construction of vertex amplitudes for
spinfoam models. In this Section, we study how the coherent tetrahedron
can be decomposed in the conventional basis of virtual links, which
we fix by choosing to add $\vec J_1$ and $\vec J_2$ first.

The coefficients entering \Ref{n0} can be studied with the same techniques used for the norms.
Introducing the shorthand notation $\ket{i}\equiv\ket{j_i, \n_i}$, in the 4-valent case we have
\equ\label{dec}
|\bra{j_1\ldots j_4, j_{12}} \underline{j}, \underline{\n} \, \ra_0|^2 =
\int dh \int dg \; d_{j_{12}} \; \chi^{(j_{12})}(g) \; \bra{12}hg\ket{12}\,\bra{34}h\ket{34}.
\nequ
As it can be verified directly,  $d_{j_{12}} \; \chi^{(j_{12})}(g)$ projects on the intertwiner state
$\ket{j_1\ldots j_4, j_{12}}$.
To study the asymptotics, it is convenient to introduce an auxiliary unit vector $\n$, writing
the character in the basis of coherent states,
\equ
\chi^{(j_{12})}(g) = d_{j_{12}} \int d^2\n \; \bra{j_{12},\n} g \ket{j_{12}, \n}.
\nequ
As in the previous Section, we parametrize the group elements as $h = p_\eta + i\vec p\cdot \vec\si$,
$g = q_{\eta'} + i\vec q\cdot \vec\si$.
We have
\equ\label{gh}
 gh  = p_\eta \, q_{\eta'} - \vec p \cdot \vec q +
i \, \Big( p_\eta \, \vec q + q_{\eta'} \, \vec p - \vec p \w \vec q \, \Big) \cdot \vec \si.
\nequ
Using this and \Ref{cis}, and rescaling again $j_i \mapsto \lambda \, k_i$, we rewrite \Ref{dec} as
\equ
|\bra{j_1\ldots j_4, j_{12}} \underline{j}, \underline{\n} \, \ra_0|^2 =
\f{d_{j_{12}}{}^2}{(2\pi^2)^2}\sum_{\eta,\eta'=\pm} \int_{B_\eta} \f{d^3 \vec{p}}{\sqrt{1-\vec{p} \,{}^2}}
\int_{B_{\eta'}} \f{d^3 \vec{q}}{\sqrt{1-\vec{q} \,{}^2}} \int d^2\n
\; e^{\lambda S(\vec p, \vec q \, )}
\nequ
with
\eqa
e^{S(\vec p, \vec q \, )} &=&
(q_{\eta'} + i \vec q \cdot \n)^{2 k_{12}}
\prod_{i=1}^2 \Big( p_\eta \, q_{\eta'} - \vec p \cdot \vec q +
i \, (p_\eta \, \vec q + q_{\eta'} \, \vec p - \vec p \w \vec q \; ) \cdot \n_i\Big)^{2 k_i}
 \no
&& 
\times
\prod_{i=3}^4 (p_\eta + i \vec p \cdot \n_i)^{2 k_i}.
\neqa
From the analysis of the previous Sections, we expect the asymptotics
of these coefficients to be dominated by $g$ and $h$ close to the identity.
Expanding $S$ around $\vec p = \vec q=0$ and denoting $\underline{r} = (\vec p, \vec q\, )$, we have
\equ
S(\underline{r} ) =2\, i \, \widetilde{{N}}\cdot \underline{r}-
\underline{r} \cdot \widetilde{H} \underline{r} + o(r^3),
\nequ
where $\widetilde{{N}}$ is the following 6-dimensional vector,
\equ
\widetilde{{N}}= \left( \begin{array}{cc}
\vec N  \\ k_1 \, \n_1 + k_2 \, \n_2 + k_{12} \, \n
\end{array}\right).
\nequ
The Hessian matrix has the following structure,
\equ
\widetilde{H} = \left( \begin{array}{cc}
H & F \\ F^{\scr T} & H'
\end{array}\right),
\nequ
with $H$ the same as in \Ref{H}, and
\eqa
F_{ab} &=& k_1 P_{ab}^{1} + k_2 P_{ab}^{2}  - i \,
\eta \, \eta' \, \eps_{abc} \, \Big(k_1 \, (\n_1)_c + k_2 \, (\n_2)_c \Big),
\\
H'_{ab} &=& k_1 P_{ab}^{1} + k_2  P_{ab}^{2} + k_{12}  P_{ab}.
\neqa
The antisymmetric part of $F$ comes from the $\vec p \w \vec q$ term in \Ref{gh}, so from the
non-abelian nature of $\SU(2)$. Notice that the only imaginary term in $\widetilde{H}$ is
proportional to $\eta \, \eta'$, thus when we perform the sums in the Gaussian approximation, we
are simply going to get (four times) the real part of a single Gaussian integration.

\begin{figure}[ht]
\includegraphics[width=6cm]{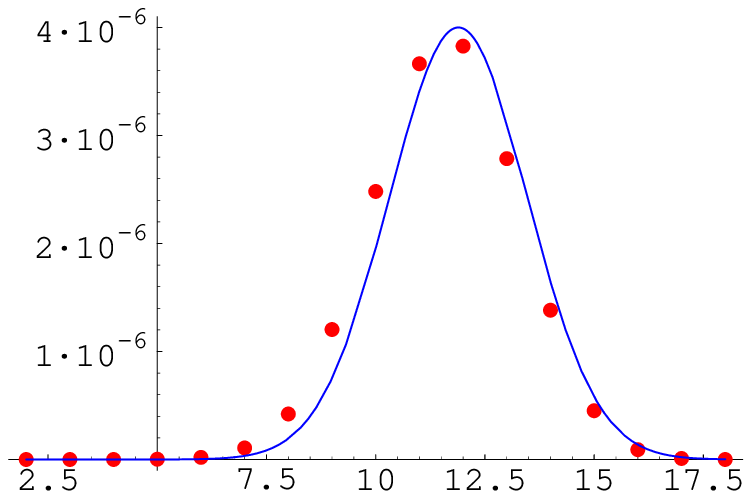} \hspace{1cm}
\includegraphics[width=7cm]{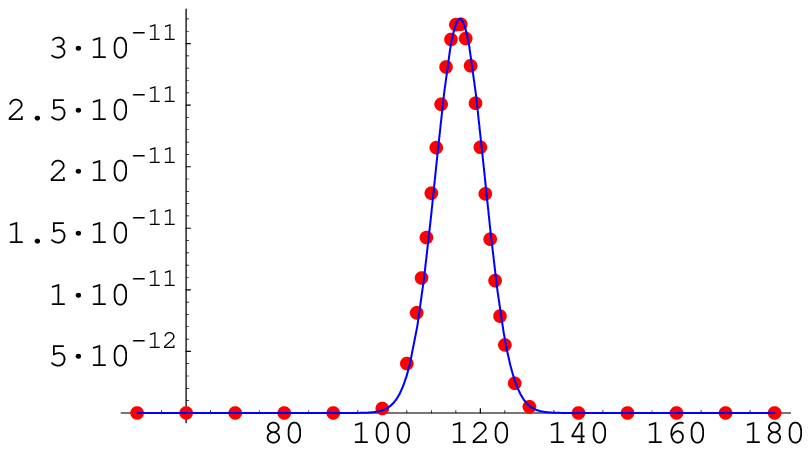}
\caption{\footnotesize  Plots of \Ref{prob} for the equilateral configuration $j_i = j_0$ $\forall i$.
The dots represent the exact numerical evaluation, whereas the line is the Gaussian
\Ref{gausscoeff}. The left panel shows the case $j_0=10$, whereas the right panel shows the case
$j_0=100$. In the small spin case, the Gaussian approximation is already capturing the right
behavior, and it becomes very accurate in the large spin case. \label{totti} }
\end{figure}

The asymptotics are given by
\equ\label{prob}
|\bra{j_1\ldots j_4, j_{12}} \underline{j}, \underline{\n} \, \ra_0|^2 \simeq
\f{d_j{}^2}{\pi \, \lambda^3} \sum_{\eta, \eta' =\pm}
\int d^2\n \ \f1{\sqrt{\det \widetilde{H}}}\
\Big(e^{- \widetilde{N} \cdot \widetilde{H}^{-1} \widetilde{N}} \Big)^\lambda,
\nequ
where $\det\wtl{H}$ can expressed in term of $3\times3$ determinants as
$\det H \det(G-F^{\scr T}H^{-1}F)$. This integral can be studied numerically. For fixed $j_i, \
\n_i$, it represents the probability of the eigenstate $j_{12}$ as a function of the $\n_i$'s. For
closed configurations, we expect this to be a Gaussian peaked on the semiclassical value computed
from the $\n_i$'s. Let us consider for simplicity the equilateral case. In this case, we expect
$j_{12}$ to be peaked around $\overline{j}$ such that $\overline{j} (\overline{j}+1) = \mean{(\vec
J_1 + \vec J_2)^2} = 2j_0 (\f23 j_0 + 1)$, which gives $\overline{j} = \sqrt{2j_0 (\f23 j_0 +
1)+\f14}-\f12$. As we show in Fig.\ref{totti}, for large spin \Ref{prob} is indeed approximated by
the Gaussian
\equ\label{gausscoeff}
p(j_{12}) = N(j_0) \exp\left\{-\f{(j_{12} - \overline{j}\,)^2}{\si} \right\},
\nequ
where $N(j_0)$ is the normalization.
Confronted with the numerics, we fix $\si = j_0/2$.
This semiclassical property is manifestly preserved by changing the pairing,
namely the choice of basis. For any choice of $j_{ik}$ the probability \Ref{prob} is peaked on
$j_{ik}(j_{ik}+1) \sim (j_i \, \n_i + j_k \, \n_k)^2$  with vanishing relative uncertainty.\footnote{An
alternative definition of coherent states, considered in \cite{tesi}, is to require the minimization of the uncertainty
$\Delta \vec J_i\cdot \vec J_k \,\Delta \vec J_i\cdot \vec J_l \geq
\f12 |\mean{[\Delta \vec J_i\cdot \vec J_k, \Delta \vec J_i\cdot \vec J_k ]}| \equiv
\f12 |\mean{\eps_{abc} \,J_i^a \,J_k^b \,J_l^c}| \sim O(j^3)$. The states constructed here
do not stricly minimize this, even if the do satisfy the non trivial condition
$\Delta \vec J_i\cdot \vec J_k \,\Delta \vec J_i\cdot \vec J_l \sim O(j^3)$ which guarantees
the vanishing relative uncertainties.
}

Analogous results hold for arbitrary closed configurations. This shows in a concrete way
in which sense $\ket{\underline{j}, \underline{\n} \, }_0$ represents a semiclassical state for a quantum tetrahedron, and more in general for a $V$-simplex dual to a node of valence $V$.

At this point, it is useful to compare our construction of the semiclassical tetrahedron with the
one introduced in \cite{semi}.
Both states have the property that for any $i$ and $k$, $\mean{\vec J_i \cdot \vec J_k}$ is peaked
around a given semiclassical value with vanishing relative uncertainty in the large spin limit.
The uncertainties are slightly different, as can be seen by direct comparison in the equilateral case.
From equation (41) of \cite{semi} (with the appropriate redefinition of $j_0$) we read
$\si=4j_0/3\sqrt 3$ which is roughly one and a half times the spread of \Ref{gausscoeff}.
The main advantage of the states introduced here lies in the fact that they allow the construction of
a wider framework.
The construction in \cite{semi} is only valid for a tetrahedron, and tailored to studying the
large spin limit only. The framework considered here, on the other hand, applies to any $V$-simplex,
and it is well--defined for any spin. In particular, the key feature of the coherent intertwiners considered here
is that they provided an overcomplete basis in $\H_0$, which makes them useful to study the dynamics,
as we described in Section \ref{SecVertex}.

\section{Towards the quantum gravity amplitude}\label{SecQG}

Let us finally discuss how the coherent state technology presented here can be used to define the
dynamics of quantum GR, starting from the spinfoam model \Ref{Z2}. Spinfoam quantization of GR usually relies
on reformulating GR as a constrained BF theory with an action of the following type \cite{Plebanski,
Reisenberger,Perez},
\equ\label{plebanski}
S_{\rm GR}(B_{\mu\nu}, \om_\mu)  = \int {\rm Tr} \, B\w F(\om) + {\cal C}(B_{\mu\nu}).
\nequ
$\om_\mu$ is a connection valued on a given Lie algebra (for euclidean signature, $\su(2)$ or
$\spin(4)$), and $B_{\mu\nu}$ is a bivector field (or two-form) with values in the same Lie
algebra. The term ${\cal C}(B)$ includes polynomial constraints, reducing topological BF to
GR.\footnote{The logic of this reduction is the following. The initial BF action describes a
topological theory with no local degree of freedom, with the field $B$ a Lagrange multiplier
enforcing that the connection is flat, $F(\om)=0$. The term ${\cal C}(B)$ then constrains the
Lagrange multiplier, thus enlarging the phase space. It breaks the topological invariance
(translation symmetry of $B$) and introduces non-trivial local degrees of freedom.} Typically, it
gives the set of second class constraints expressing $B$ in terms of the tetrad field (or vierbein)
and leading to GR in the first order formalism (for a detailed canonical analysis of
\Ref{plebanski}, see \cite{horrible}).

The BF theory can be quantized as described in Section \ref{SecVertex}, discretizing
the spacetime manifold with a Regge triangulation (or more generally a cellular decomposition), and then
evaluating the partition function \Ref{Z}, where the variables are the
representations $j_t$ and intertwiners $i_\tau$.
The natural extension of this procedure to quantize GR with action \Ref{plebanski} would be to
discretize the constraints ${\cal C}(B)$ and include them in the computation of the discretized
path integral (see e.g. \cite{worldsheet,actionprinciple}). Nevertheless, on the grounds
of geometric quantization (see e.g. \cite{BarrettC}), one usually shortcuts this computation and directly
attempts to translate the discretized constraints ${\cal C}(B)$ as constraints on the variables
$j_t$ and $i_\tau$ by assuming that the variables $B$ are represented in the spinfoam as the
generators of the considered Lie group, as in \Ref{BJ} for $\SU(2)$.
This procedure is usually referred to as imposing the constraints at the quantum level,
and it leads to a unique choice of intertwiner and to the Barrett-Crane model \cite{BarrettC,unique}.
However, the quantum constraints obtained in this way are fairly strong, and a number of authors
have raised various doubts on this procedure, such as:
the rewriting of the constraints in purely algebraic terms (representations and intertwiners)
partially hides their geometric meaning, and thus the interpretation of the model;
the unique intertwiner does not seem to be compatible with the
data on the boundary spin network as given in LQG; more crucially, the quantum constraints do not commute
with each other and generate (by commutator) higher order constraints which do not seem to have any classical
equivalent.

A plausible explanation of these difficulties is that the BC model imposes the constraints too
strongly, and thus does not have enough degrees of freedom to describe quantum GR. Then a way out
would be to loosen the implementation of the constraints, by requiring them to hold only on average
(vanishing expectation value) with minimal uncertainty. This seems the natural procedure when
dealing with non-commuting constraints.\footnote{A possible strategy to impose strongly second
class constraints is discussed in \cite{Ashtekar}.} It has been suggested for instance in
\cite{worldsheet}, where both possibilities are discussed, (i) to impose the constraints exactly by
inserting the $\delta$-distribution $\delta({\cal C}(B))$ in the path integral, or (ii) to
regularize the $\d$-distribution by inserting its Gaussian approximation $\exp(-{\cal
C}(B)^2/\xi^2)$, with $\xi$ a free parameter. Notice that the Gaussian insertion amounts to impose
the constraints approximatively. More precisely, since the $B$ field represents the geometry and
large values of $B$ corresponds to large (length) scales, using the Gaussian means imposing the
constraints at the semiclassical level while allowing slight deviations at smaller scales.

In the same spirit, here we argue that the partition function \Ref{Z2} for the quantum BF
theory in terms of coherent states offers a natural way to impose the constraints on average.
The key is that the partition function \Ref{Z2} provides a quantization of BF theory where the $B$ field
is represented not through generators of the group,
but as representation  and intertwiner labels, such as $j_t$ and $\n_t$ (see
\Ref{B} or \Ref{B1}). This representation allows to write directly the constraints ${\cal C}(B)$
as a sum of local contributions ${\cal C}_\si(j_t\,\n_{t,\tau})$, in terms of the dynamical
variables and with a clear geometric interpretation. In particular, the semiclassical dynamics
becomes transparent. As we understand from the analysis reported in the previous Sections, the
vertex is dominated in the large spin limit by semiclassical states satisfying the closure
condition for each tetrahedron and the relation \Ref{normals} between adjacent tetrahedra in the
same 4-simplex. On these states the variables $j_t$ and $\n_t$ give classical values with an
(almost minimal) uncertainty decreasing as the $j_t$'s increase. Therefore, imposing
${\cal C}_\si(j_t\,\n_{t,\tau})=0$ amounts to impose the classical constraints only on average
with a small uncertainty vanishing in the large spin limit, namely semiclassically.
If we use the $j_t$'s and $\n_t$'s to construct a Regge geometry, we expect that the
role of the constraints is to generate deficit angles when we glue together various 4-simplices,
thus allowing the geometry to be curved. We postpone the precise analysis of the implementation of
the constraints for further work.

\section{Conclusions}\label{SecConcl}

The standard intertwiner basis which leads to
BF theory with the vertex amplitude given by the $\{15j\}$ symbol does not appear to be the most suitable
one to study the semiclassical geometry of BF theory. Furthermore,
it makes it hard to understand the quantum structure of the constraints reducing
BF to GR, and possibly hides the correct way to implement them.

Here we considered a basis constructed out of $\SU(2)$ coherent states. We defined
non-normalized coherent intertwiners, and studied their norm as a function of the geometric
configuration described. For each configuration, the norm is an integral over $\SU(2)$ that can be solved exactly
as described in the Appendix. Yet a saddle point evaluation of the leading order of this
integral in the large spin limit proves a very accurate approximation even for small spins,
and shows very neatly that the norm is exponentially maximized by the states admitting a
semiclassical interpretation, namely the ones whose quantum numbers
can be interpreted as vectors $j_i \, \n_i$ describing the classical
discrete geometry of a $V$-simplex. Thanks to this crucial result, the semiclassical states will dominate
the evaluation of quantum correlations.

Using these coherent intertwiners we rewrote the BF partition function with a new vertex
amplitude, given in \Ref{vertex}, where the discrete $B_t(\tau)$ variables are interpreted in terms of
the vectors $j_t \, \n_{t,\tau}$, thus retaining the original vectorial nature of the $B$ field. We
expect this reformulation of the BF spinfoam amplitudes to improve the geometric interpretation of
the theory and in particular our understanding of what should be the proper way to implement the
constraints reducing it to GR.

We hope that the developments presented here will contribute to the promising recent
advances in understanding the low energy limit of LQG, such as
the ones on the graviton propagator \cite{grav2,RovelliProp,grav3,Io},
on the emergence of effective actions for matter \cite{etera}, or on applications to cosmology \cite{cosmo}
and black hole physics \cite{bh}.

\section*{Acknowledgments}

Research at Perimeter Institute for Theoretical Physics is supported in
part by the Government of Canada through NSERC and by the Province of
Ontario through MRI.

\appendix

\section{Coherent States for $\SU(2)$: a brief review}
\label{AppCS}

$\SU(2)$ coherent states minimize the ($\SU(2)$
invariant) uncertainty $\Delta\equiv\,|\la\vec{J}^2\ra-\la\vec{J}\ra^2|$
in the direction of the angular momentum \cite{Perelomov}.
A simple calculation shows that on a basis state $|j,m\ra$ the uncertainty
$
\Delta(j,m)=j(j+1)-m^2
$
is minimal when $m=j$. The maximal weight vectors $\ket{j,j}$ are thus coherent states
for arbitrary choice of spin $j$ and angular momentum axis $J_z$.
Starting from the highest weight, an infinite set of coherent states
on the sphere $\SU(2)/\U(1)\sim {S}^2$ are constructed through the group action,
$
\ket{j,\n} =g(\hat{n})\ket{j,j},
$
where $\n$ is a unit vector defining a direction on the
sphere ${S}^2$ and $g(\hat{n})$ a $\SU(2)$ group element
rotating the direction $\hat{z}\equiv (0,0,1)$ into the direction
$\hat{n}$.
Explicitly, taking
$\hat{n}=(\sin\theta\cos\phi,\sin\theta\sin\phi,\cos\theta)$, we
choose $g(\hat{n})\equiv \exp\{i\th \, \hat{m}\cdot\vec{J}\}$ where
$\hat{m}\equiv(\sin\phi,-\cos\phi,0)$ is a unit vector orthogonal
to both the directions $\hat{z}$ and $\hat{n}$.
Just as $\ket{j,j}$ has direction $z$ with minimal uncertainty, $\ket{j, \n}$ has direction $\n$
with minimal uncertainty, as can be verified explicitly using the formula reported in the next Appendix.

A coherent state can be decomposed in the usual basis as
\equ\label{coher}
\ket{j,\n} = \sum_{m=-j}^j a_m(\n) \ket{j, m},
\nequ
where
$$
a_m(\n) = \sqrt{\f{(2j)!}{(j-m)! (j+m)!}}\f{\zeta^{j-m}}{(1+|\zeta|^2)^j}, \qquad
\zeta=\tan\f\theta2 e^{-i\phi}.
$$

These states are normalized but not orthogonal, the scalar product between two of them being
$$
\la j,\hat{n}_1|j,\hat{n}_2\ra=\left(\f{1+\hat{n}_1\cdot\hat{n}_2}{2}\right)^j\,
e^{ijA(z,\hat{n}_1,\hat{n}_2)},
\qquad
A(z,\hat{n}_1,\hat{n}_2) = -\f i2 \ln \left(\f{1+\xi_1\overline{\xi_2}}{1+\overline{\xi_1}\xi_2}\right).
$$
Here $\xi=\f\theta2 e^{-i\phi}$.
Notice that $A$ is the area of the geodesic
triangle on the sphere $\ss^2$ by the north pole direction z
and the two unit vectors $\hat{n}_1$ and $\hat{n}_2$.

Consequently they provide an overcomplete basis for the irreps $j$, and the
resolution of the identity can be written as
${\mathbbm 1}_j = d_j \int d^2\n \, \ket{j, \n} \bra{j, \n}$, with $d^2\n$
the normalized measure on $\ss^2$.

The explicit coefficients entering \Ref{n0} can be found using \Ref{coher}
to decompose $\ket{\underline{j}, \underline{\n} \, }_0 $ into the conventional basis of $\H_0$,
\eqa\label{n01}
\H_0 \ni \quad \ket{\underline{j}, \underline{\n} \, }_0 
= \sum_{m_1\ldots m_V} \prod_{i=1}^V a_{m_i}(\n_i)  \sum_{i_1\ldots i_{V-3}}
C^{i_1\ldots i_{V-3}}_{m_1\ldots  m_V} \ \ket{j_1 \ldots j_V, i_1 \ldots i_{V-3}},
\neqa
where we have introduced the (generalized) Clebsch-Gordan coefficients $C_{m_1\ldots
m_V}^{i_1\ldots i_{V-3}}$ (see \cite{Brink}). These are defined\footnote{Here we loosely use the term
Cebsch-Gordan coefficients, to refer to invariant tensors.
These can differ in phase and normalization from other definitions found in the literature.}
by the integration of irrep matrices,
\equ
\int dg \, \prod_{i=1}^V D^{(j_i)}_{m_in_i}(g)
=\sum_{i_1\ldots i_{V-3}} \,
\overline{C^{i_1\ldots i_{V-3}}_{m_1\ldots  m_V}} \ C^{i_1\ldots i_{V-3}}_{n_1\ldots  n_V}.
\nequ
Using the recoupling theory, thes generalized coefficients
and can always be decomposed into sums of products of conventional (3-valent) Clebsch-Gordan coefficients.

From \Ref{n01} we immediately read the coefficients entering \Ref{n0}.

\section{Useful formulas}\label{formulas}

We report here the results used in Section \ref{SecCI}.
The calculations can be done using the explicit expression \Ref{coher}, however
it is usually easier to exploit the fact that
$$
\bra{j, \n} J_a \ket{j, \n} = \bra{j, j} J_a' \ket{j, j}
$$
where $J_a' = g(\n)^{-1} \, J_a \, g(\n)$ is the rotated generator.

We compute the following averages of the $\SU(2)$ generators on the coherent states:
\equ\label{J}
\bra{j, \n} J_a \ket{j, \n} = j \, n_a,
\qquad
\bra{j, \n} J_a^2 \ket{j, \n} = \f j2+j(j-\f12) \, n_a^2.
\nequ
From this it is easy to check that
\equ\nonumber
\Delta^2\,\equiv\,\bra{j, \n} \vec{J}\,{}^2 \ket{j, \n}-
\bra{j, \n} \vec{J}\, \ket{j, \n}\bra{j, \n} \vec{J}\, \ket{j, \n}
\,=\, j.
\nequ
Further calculations give
\equ\nonumber
\bra{j, \n} J_a \, J_b \ket{j, \n} = \f j2\,(\delta_{ab}+i \eps_{abc}\,n_c)+ j(j-\f12) \, n_a \, n_b,
\nequ
\eqa\nonumber
\bra{12} (\vec J_1+ \vec J_2)^2 \ket{12} =
(j_1\,\n_1 + j_2\n_2)^2+j_1+j_2,
\neqa
(here notice that the $j_1+j_2$ term comes from the uncertainty $\Delta^2$ computed above), and
\eqa
\bra{12} (\vec J_1+ \vec J_2)^4 \ket{12} &=&
\big( (j_1\,\n_1 + j_2\n_2)^2+j_1+j_2 \big)^2 + \no\nonumber &&
+ 2 j_1 j_2 \Big[ \Big(j_1+j_2-\f12\Big)\Big{(}1-(\n_1 \cdot \n_2)^2\Big{)} +
\left(1- 3 \, \n_1 \cdot \n_2\right) \Big].
\neqa
Therefore the uncertainty $\la(\vec J_1+ \vec J_2)^4\ra-\la(\vec J_1+ \vec J_2)^2\ra^2$ depends
explicitly on the value of the angle $\cos\theta_{12}=\n_1\cdot\n_2$. For large $j_1,j_2$,
it is for aligned vectors and maximal for $\n_1$ and $\n_2$ orthogonal.

Proceeding as above, one can show \Ref{ev} for any observable $\hat{\cal O}(\vec J_i \,)$.

Let us now report other results used in the main body of the paper.
\begin{itemize}
\item The matrix element entering the bivalent norm \Ref{abidal} can be computed directly using
the conventional parametrization
\equ\label{gi}
g(\n_i) \equiv g(\th_i, \hat m_i) = \cos\th_i \, \mathbbm{1}+ i \, \sin\th_i \, \hat m_i\cdot \vec\si,
\nequ
we have
\eqa
&& \hspace{-1.8cm}| \bra{+} g_1^{-1} g_2 \ket{-} |^{2} =
\cos^2\th_1+\cos^2\th_2 -2 \cos\th_1 \cos\th_2 \, (\cos\th_1 \cos\th_2 +
\sin\th_1 \sin\th_2 \; \hat{m}_1\cdot \hat{m}_2) = \no \no
&& \hspace{0.9cm}= \f{1-\n_1\cdot \n_2}2 = 1-\f14\, (\n_1+\n_2)^2.
\neqa

\item
In Section \ref{SecCI} we also made use of the adjoint action of the group on itself.
Using \Ref{g} and \Ref{gi}, we have
$$
g(\n)^{-1} \, h \, g(\n) = \cos\ga + i\, \sin\ga \, \hat u' \cdot \vec\si,
\qquad
\hat u' = g(\n)^{-1}\, \hat{u}.
$$
By definition $g(\n)$ rotates the north pole vector $(0,0,1)$ to the direction $\hat{n}$,
thus $(\hat u')_3=(g(\n)^{-1}\, \hat{u})_3=\hat{u}\cdot\hat{n}$, as can be
explicitly checked writing the components of the rotated vector $\hat u'$:
$$
\hat u' \,=
\cos\th \, \hat u - \sin\th \, \hat u\w \hat m +
2 \,\sin^2\f\th2 \, (\hat u\cdot \hat m) \, \hat m,
$$
$$
(\hat u')_3= \cos\th \, \hat u_3 - \sin\th \, (\hat u\w \hat m)_3 =
\cos\th \, \hat u_3 + \sin\th \, (\hat u_1 \cos\phi + \hat u_2 \sin\phi) \,=\, \hat u \cdot \n.
$$
From this we immediately derive \Ref{cis}.

\item
A crucial result concerns the extension of \Ref{ev} to expectation values in $\H_0$. To prove this, one
can compute
$$
\bra{j, \n} h \, \vec J \, \ket{j, \n}\,=\,
j\big{[}\cos\ga \, \hat{n}+i \, \sin\ga \, \hat{u}+\sin\ga \, \hat{u}\w \hat{n}\big{]}
\,(\cos\ga+i\, \sin\ga \, \hat{u}\cdot\hat{n})^{2j-1}.
$$

As we see, the presence of the group element $h$ complicates the expectation values of the $\SU(2)$
generators (compare the above with \Ref{J}).
However, we know from the analysis in Section \ref{SecCI} that the integration over $h$ peaks
the group elements on the identity $\ga = 0$ in the large spin limit,
thus
$$
\int dh \, \bra{j, \n} h \, \vec J \, \ket{j, \n} \simeq j \, \n,
$$
and all the results reported above extend naturally to $\H_0$ in the large spin limit.

\end{itemize}

\section{Evaluating the norm using spherical integrals}\label{AppSpher}

In Section \ref{SecCI} we wrote the norm \Ref{normcos} using $\vec p$ to parametrize $\SU(2)$.
This choice was convenient to study the saddle point approximation.
On the other hand, to evaluate the norm exactly, it is more useful the standard parametrization
\Ref{g} in terms of a rotation angle $\ga$ and its rotation axis $\hat{u}$,
$$
f(\hat{n}_i)\,=\,
\int dh\, \bra{j_i,\hat{n}_i}h^{\otimes V}\ket{j_i,\hat{n}_i}
\,=\,
\f{2}{\pi}\int_0^\pi d\ga \,  \sin^2\ga \, \int_{\ss^2}{d^2\hat{u}} \
\prod_{i=1}^V (\cos\ga+i\sin\ga \ \hat{u}\cdot\n_i)^{2j_i}.
$$
We can expand all the terms and deal separately with the integrals over $\gamma$ and the spherical
integrals over $\hat{u}$. Using the binomial expansion and  $J\,\equiv\,\sum_i j_i$, the
$d\ga$ integrals are of the following type,
\equ
\I(J,k)\,\equiv\,\f2\pi\int d\ga\, (\sin\ga)^{2+2k}(\cos\ga)^{2J-2k}\,=\,
\f{(2J)! \, (2k+1)!! \, (2J-2k-1)!!}{2^{2J} \, (J+1) \, (J!)^2 \,(2J-1)!!},
\nequ
with $k \leq J$.

The next step is to compute the spherical integrals. One can show that:
\begin{eqnarray*}
\int d^2\hat{u} \ \prod_{i=1}^2(\hat{u}\cdot\n_i)&=&
\f{1}{3}\,\hat{n}_1\cdot\hat{n}_2, \\
\int d^2\hat{u} \ \prod_{i=1}^4(\hat{u}\cdot\n_i)
&=&
\f{1}{3\times 5}\,
\big{[}(\n_1\cdot\n_2)(\n_3\cdot\n_4)+(\n_1\cdot\n_3)(\n_2\cdot\n_4)+(\n_1\cdot\n_4)(\n_2\cdot\n_3)\big{]}.
\end{eqnarray*}
This can be generalized to even polynomials of arbitrary high order in $\hat{u}$.
A generic term $\int \prod_i^{2N} (\hat{u}\cdot \n_i)$ will have a prefactor
$(2N+1)!!$ and a sum over the $(N-1)!!$ possible pairings of the vectors $\hat{n}_i$ with each other.

Using these formulas one can compute exactly the coherent intertwiner norm $f(\hat{n}_i)$ as a
finite sum of polynomials of the $\hat{n}_i$.

The situation is particularly simple in the degenerate case, when all the $\n_i$ are aligned.
In this case, discussed in Section \ref{SecDeg}, the spherical integrals are trivial, and we have
to evaluate simply
\eqa\label{deg}
f(N, \eta_i \hat{z}) &=&
\f1\pi \int d\beta \, d\ga \, \sin\beta \, \sin^2\ga \, (1-\sin^2\beta \, \sin^2\ga)^N =
\\\nonumber &=&
\f1{\pi} \sum_{k=0}^N (-1)^k
\left(\begin{array}{cc} N \\ k \end{array}\right) \int d\beta \, (\sin\beta)^{2k+1} \int d\ga\, (\sin\ga)^{2k+2} =
\sum_{k=0}^N \f{(-1)^k}{k+1} \left(\begin{array}{cc} N \\ k \end{array}\right),
\neqa
where in the last step we used
\equ\nonumber
\int d\beta \, (\sin\beta)^{2k+1} = \sqrt\pi \, \f{\Gamma(k+\f32)}{\Gamma(k+2)},
\qquad
\int d\ga\, (\sin\ga)^{2k+2} = \sqrt\pi \, \f{\Gamma(k+1)}{\Gamma(k+\f32)}.
\nequ
This sum in \Ref{deg} can be straightforwardly evaluated with the substitution $t=s+1$, to obtain
\equ
f(N, \eta_i \, \hat{z}) =  \f1{N+1} \, \sum_{t=1}^{N+1}
\f{{(-1)^{t-1}} \, (N+1)!}{t! \, (N+1-t)!} = -\f1{N+1}\Big((1-1)^{N+1} -1 \Big)
\equiv \f1{N+1}.
\nequ
This result was used to evaluate \Ref{mb1}.

\section{Numerics}\label{AppPlots}
All the calculations of this paper have been supported by numerical simulations,
performed with Mathematica$^{\scr TM}$. In this Appendix
we report some examples to illustrate the numerical support to the saddle point analysis of
Sections \ref{SecSaddle} and \ref{SecNC}.

\begin{figure}[ht]
\includegraphics[width=6cm]{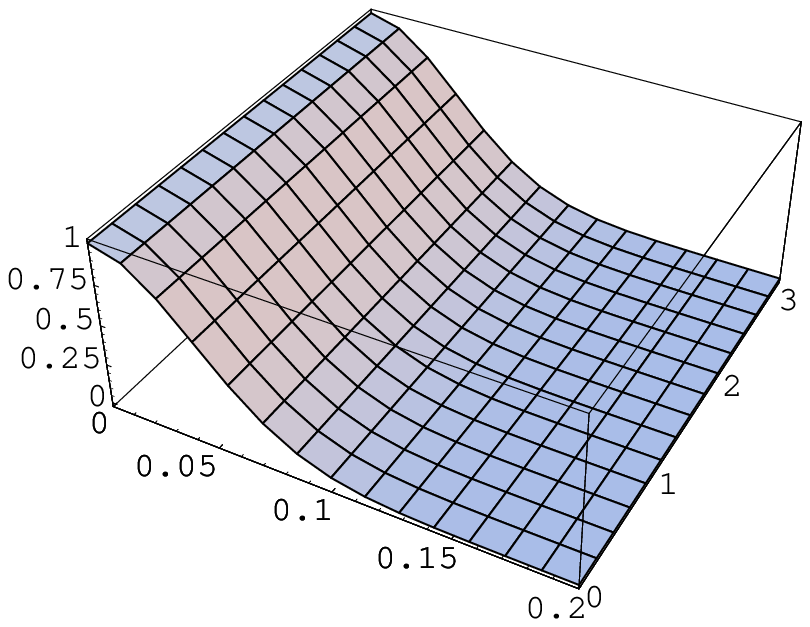} \hspace{1cm}
\includegraphics[width=6cm]{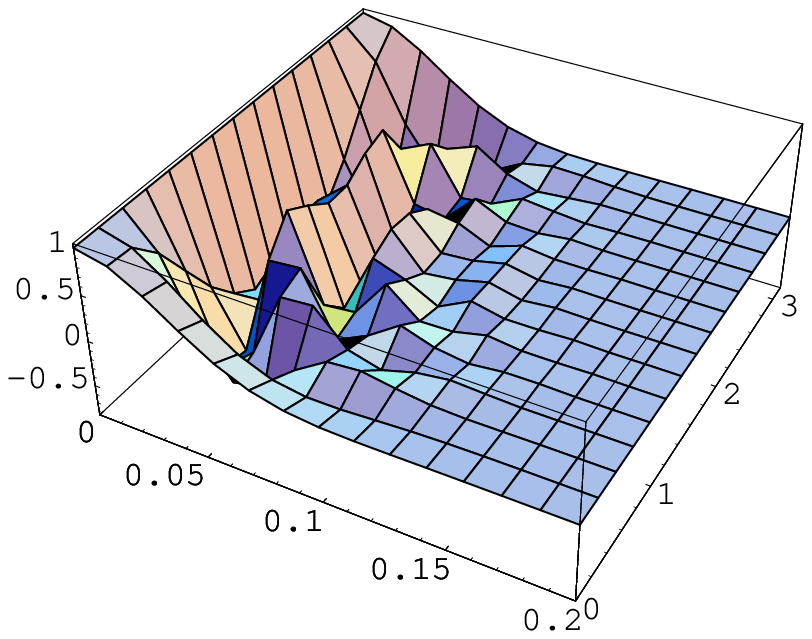}
\caption{\footnotesize The real part of the
integrand for $V=4$ as a function of $\ga\in [0,0.2]$ (symmetric at $\pi/2$) and $\beta\in [0,\pi]$,
for fixed $\alpha$. Left panel: the equilateral case, at $j_i=100$ for all $i$.
Right panel: a generic open configuration, with different spins but all of order 100.
\label{anelka}}
\end{figure}

The first thing we show is the absence of saddle points when the closure is not satisfied.
Consider the integrand in \Ref{normcos}, with the parametrization
\Ref{g} where $\hat u = (\cos\alpha \sin\beta, \sin\alpha \sin\beta, \cos\beta)$.
To give a clear picture of the integrand, we fix an arbitrary value of $\alpha$
and show a 3d plot in only $\beta$ and $\ga$. 
We choose the simplest case, the equilateral tetrahedron ($V=4$) with $j_i=100 \ \forall i$.
This is shown on the left panel of Fig.\ref{anelka}.
It is clearly visible that $\ga = 0$ is a saddle point (the degeneracy in $\beta$ is an
artifact of the polar coordinates for $\ss^3$), and that the integrand is quickly suppressed away from it.
On the right panel, we have again $V=4$, but this time we picked a random non closed configuration,
with the four different spins still of order 100, but not equal. First of all, one can notice that
even if $\ga = 0$ still maximizes the integrand, it is not anymore a saddle point.
Second, notice that the integrand has now also negative values, which contribute to dump the integral in
this non closed case.

\begin{figure}[ht]
\includegraphics[width=6cm]{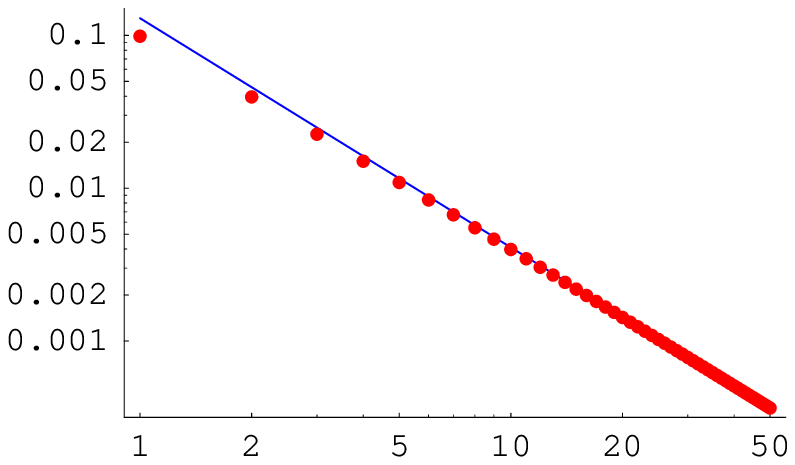} \hspace{1cm}
\includegraphics[width=6cm]{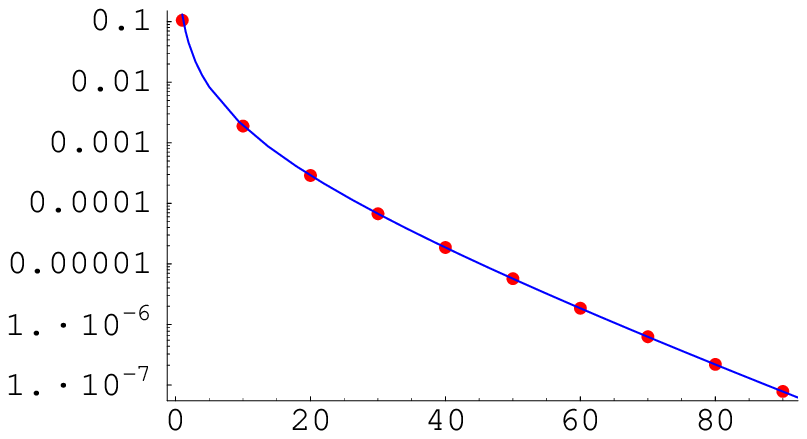}
\caption{\footnotesize Bilogaritmic plots.
\emph{Left panel:} the dots are the numerical evaluation of the exact norm
\Ref{norm} for the equilateral tetrahedron ($j_i=j$ for all $i$), for different values of the spin $j$.
The line is the analytic calculation of the leading order \Ref{saddleclosed}.
\emph{Right panel:} an open case with $j_i=j$ for all $i$ but the normals not closing.
The line is the analytic result \Ref{nonclosed}.\label{cisse}
}
\end{figure}

Next, we consider the Gaussian evaluation of the norm. In Fig.\ref{cisse} we compare
the exact numerical evaluations of \Ref{normcos} with the analytic calculations.
On the left panel, we consider a closed configuration, for simplicity given by
the equilateral tetrahedron, namely $V=4$ and $j_i=j$ $\forall i$.
The dots represent the numerical evaluations for different values of $j$, whereas
the line is the analytic result \Ref{saddleclosed}. As one can see, the agreement is very good
also at small spins. By direct reading of the numerics, one sees that the
analytic approximation matches to one digit the exact result already at $j\sim 1$,
and that at $j\sim 100$ the matching is to three digits. The situation is
very similar for arbitrary configurations.
On the right panel we consider a non closed configuration with $j_i=j$ $\forall i$.
The dots are again the numeric
evaluations, and the line the analytic result \Ref{nonclosed}.
Notice that the agreement is still very good even at small spins.
As expected, the norm is exponentially smaller than in the closed configuration.
In the open case, the integral converges poorly, for this reason we have less points in the plot.

The results reported above show how accurate the saddle point approximation is.


\end{document}